\documentclass[twocolumn]{svjour3}
\usepackage{graphicx}
\usepackage{amsmath,amssymb}

\journalname{Eur. Phys. J. C}

\begin{document}

\title{Investigation of $\Xi_c^0$ in a chiral quark model}

\author{Xiaohuang Hu \and Yue Tan \and Jialun Ping\thanks{Corresponding author: jlping@njnu.edu.cn}}

\institute{Department of Physics and Jiangsu Key Laboratory for Numerical
Simulation of Large Scale Complex Systems, Nanjing Normal University, Nanjing 210023, P. R. China}

\titlerunning{Investigation of $\Xi_c^0$ in a chiral quark model}
\authorrunning{Hu,Tan,Ping}

\maketitle

\begin{abstract}
  Recently, three new states of $\Xi_c^0$ were observed in the invariant mass spectrum of $\Lambda^+_cK^-$ by LHCb collaboration. In this work,
  we use a chiral quark model to investigate these three exited states with the help of Gaussian expansion method both in three-quark structure
  and in five-quark structure with all possible quantum numbers $IJ^P=\frac{1}{2}(\frac{1}{2})^-$, $\frac{1}{2}(\frac{3}{2})^-$,
  $\frac{1}{2}(\frac{5}{2})^-$, $\frac{3}{2}(\frac{1}{2})^-$, $\frac{3}{2}(\frac{3}{2})^-$ and $\frac{3}{2}(\frac{5}{2})^-$ .
  The calculations shows that the 2$S$ states of $\Xi_c'(2579)^0$ and $\Xi_c(2645)^0$ are comparable to experimental results; In addition,
  the resonance states of five-quark configuration are possible candidates of these new states with negative parity by using the real scaling
  method and their decay width are also given.
\end{abstract}

\maketitle

\section{Introduction}
During recent years, many narrow $\Xi_c$ states composed of a charmed quark ($c$) a strange quark ($s$) and a light quark ($u$ or $d$) have been 
reported by Belle and BaBar Collaborations~\cite{1,2,3,4,5,6}. Very recently, LHCb Collaboration announced that three other $\Xi_c^0$ resonances 
$\Xi_c(2923)^0$, $\Xi_c(2938)^0$ and $\Xi_c(2965)^0$ have been observed in the $\Lambda^+_cK^-$ mass spectrum~\cite{7}. The mass and widths of 
the observed resonances are shown below,
\begin{eqnarray}
\Xi_c(2923)^0: && M = 2923.04 \pm 0.25 \pm 0.20 \pm 0.14 \mbox{MeV} \nonumber \\
&& \Gamma = 7.1 \pm 0.8 \pm 1.8 \mbox{MeV}  \nonumber  \nonumber \\
\Xi_c(2938)^0: && M = 2938.55 \pm 0.21 \pm 0.17 \pm 0.14 \mbox{MeV}  \nonumber \\
&& \Gamma = 10.2 \pm 0.8 \pm 1.1 \mbox{MeV}  \nonumber \\
\Xi_c(2965)^0: && M = 2964.88 \pm 0.26 \pm 0.14 \pm 0.14 \mbox{MeV}  \nonumber \\
&& \Gamma = 14.1 \pm 0.9 \pm 1.3 \mbox{MeV}  \nonumber
\end{eqnarray}
The above results have led to hot discussions while the quantum numbers of these resonances are still unknown. In Ref.~\cite{7}, LHCb Collaboration
pointed out that the $\Xi_c(2930)^0$~\cite{2,3} should be the overlap of state $\Xi_c(2923)^0$ and state $\Xi_c(2938)^0$ while the
$\Xi_c(2970)^0$~\cite{4,5,8} may be different from the state $\Xi_c(2965)^0$. In Ref.~\cite{9}, three newly observed states were studied in the QCD sum rule
and light-cone sum rules, and their results suggest that three new states can be well explained as $\Xi_c'$ baryons with quantum numbers
$\frac{1}{2}(\frac{1}{2})^-$ or $\frac{1}{2}(\frac{3}{2})^-$. In Ref.~\cite{10}, the three $\Xi_c^0$ states and their two body strong decays were evaluated
within a chiral quark model and it was found that the resonances $\Xi_c(2923)^0$ and $\Xi_c(2938)^0$ are most likely to be 1$P$ $\Xi_c'$ states with
$IJ^P=\frac{1}{2}(\frac{3}{2})^-$ while the $\Xi_c(2965)^0$ should be 1$P$ $\Xi_c'$ states with $IJ^P=\frac{1}{2}(\frac{5}{2})^-$.
In Ref.~\cite{11}, the author performed a $^{3}P_0$ model analysis which contain strong decay behaviors and suggested that the $\Xi_c(2923)^0$ and
$\Xi_c(2938)^0$ might be 1$P$ $\Xi_c'$ states and $\Xi_c(2965)^0$ can be 2$S$ $\Xi_c'$ states. In addition, lattice simulations also investigated these heavy flavored baryons~\cite{12,13,14}. Since the quantum numbers of these new states are not determined for the moment and
explanation of them as the excited states of three-quark baryon may be reasonable.

However, the possibility of the five-quark structure of these newly observed states cannot be excluded. From PDG~\cite{1}, the recognized three $\Xi_c^0$
ground states are $\Xi_c(2471)^0$ with $J^P=(\frac{1}{2})^+$ , $\Xi_c(2579)^0$ with $J^P=(\frac{1}{2})^+$ and $\Xi_c(2645)^0$ with $J^P=(\frac{3}{2})^+$.
The mass difference between the three newly observed excited states and the ground states are 320-494 MeV and it is enough to excite a light
quark-antiquark pair from the vacuum to form a pentaquark structure. So it is also reasonable for us to use five-quark structure dynamical calculation to
analyze the inner structure of the three newly observed states. In Ref.~\cite{15}, the excited $\Xi_c^0$ states were explained as molecular states of
$D\Sigma$-$D\Lambda$. In fact, several theoretical analyses and predictions described the $\Xi_c^0$ as molecular states~\cite{16,17,18,19}.

From the situation we have presented so far, different calculations give different explanations for the three newly observed states' structures.
So it is still controversial for the three states to be described as three-quark structure or as five-quark structure or even as a mixture of them.

Since the quark model was proposed by M. Gell-Mann and G. Zweig in 1964, respectively~\cite{20,21}, it has become the most common approach to study
the multiquark system as it has evolved. In this work, a constituent quark model (ChQM) is employed to investigate systematically the three-quark and
five-quark states corresponding to $\Xi_c^0$. To calculate accurately the results of each possible system, Gaussian expansion method (GEM)~\cite{22},
an accurate and universal few-body calculation method is used. Within this method,
the orbital wave functions of all relative motions of the systems are expanded by gaussians. And then after considering all possible color (color-singlet
and color-octet), spin and flavor configurations, we can completely determine the structures of the system. Finally, with the help of
``real scaling method", we can find the genuine five-quark resonances and their respective decay widths.

The paper is organized as follow. In Section II, details of a chiral quark model and calculation method are introduced. The results of the three-quark
structure are also presented in Section II. In Section III, the numerical results with analysis and discussion of five-body structure are presented.
Finally, We give a brief summary of this work in the last section.

\section{Chiral quark model and wave function}

In this paper, we employ the chiral quark model to investigate the states. The chiral quark model has become one of the most common approach to
hadron spectra, hadron-hadron interactions and multiquark states for its successful descriptions~\cite{23}. In this model, the massive constituent quarks
interact with each other through Goldstone boson exchange in addition to one-gluon exchange (OGE). Besides, the color confinement and the scalar octet
(the extension of chiral partner $\sigma$ meson) meson exchange are also introduced. More details of this model can be found in Ref.~\cite{23,24}.
The Hamiltonian of the chiral quark model is given as follows:
\begin{eqnarray}
H & = & \sum_{i=1}^{n}\left( m_i+\frac{p^2_i}{2m_i}\right)-T_{CM}+\sum_{j>i=1}^{n} V_{ij},  \\
V_{ij}  & = & V^{CON}_{ij}+V^{OGE}_{ij} +V^{\chi}_{ij}+V^s_{ij}, \\
V^{CON}_{ij} & = & \boldsymbol{\lambda}_i^c\cdot \boldsymbol{\lambda}_j^c
   \left[-a_c (1-e^{-\mu_cr_{ij}})+\Delta \right] , \\
V^{OGE}_{ij} & = & \frac{1}{4}\alpha_s \boldsymbol{\lambda}_i^c \cdot \boldsymbol{\lambda}_j^c
   \left[ \frac{1}{r_{ij}}-\frac{\boldsymbol{\sigma}_i\cdot \boldsymbol{\sigma}_j}{6m_im_j} 
   \frac{e^{-r_{ij}/r_0(\mu)}}{r_{ij}r^2_0(\mu)}\right] , \nonumber \\
   & & r_0(\mu)=\hat{r}_0/\mu,~~\alpha_{s} =\frac{\alpha_{0}}{\ln((\mu^2+\mu_{0}^2)/\Lambda_{0}^2)}.  \\
V^{\chi}_{ij} & = & v_{\pi}({{\bf r}_{ij}})\sum_{a=1}^{3}
	\lambda_i^a \lambda_j^a+v_{K}({{\bf r}_{ij}})\sum_{a=4}^{7}
	\lambda_i^a \lambda_j^a+v_{\eta}({{\bf r}_{ij}}) \nonumber \\
 & &  + [\cos\theta_{P}(\lambda_i^8 \lambda_j^8)-\sin\theta_{P}(\lambda_i^0 \lambda_j^0)] ,  \\
v^{\chi}_{ij} & = & \frac{g^2_{ch}}{4\pi}\frac{m^2_\chi}{12m_im_j}
	\frac{\Lambda^2_\chi}{\Lambda^2_\chi-m^2_\chi}m_\chi \nonumber \\
& &	\left[ Y(m_{\chi}r_{ij})-\frac{\Lambda^3_\chi}{m^3_\chi}Y (\Lambda_{\chi}r_{ij}) \right]
	(\boldsymbol{\sigma}_i \cdot \boldsymbol{\sigma}_j),  \nonumber \\
& & \chi=\pi,K,\eta , \\
V^{s}_{ij} & = & v_{\sigma}({{\bf r}_{ij}})
	\lambda_i^0 \lambda_j^0+v_{a_0}({{\bf r}_{ij}})\sum_{a=1}^{3}
	\lambda_i^a \lambda_j^a   \nonumber \\
& & +v_{\kappa}({{\bf r}_{ij}})\sum_{a=4}^{7}
	\lambda_i^a \lambda_j^a+v_{f_0}({{\bf r}_{ij}})
    \lambda_i^8 \lambda_j^8 ,  \\
v^{s}_{ij} & = & -\frac{g^2_{ch}}{4\pi} \frac{\Lambda^2_s}{\Lambda^2_s-m^2_s}m_s 
	\left[ Y(m_{s}r_{ij})-\frac{\Lambda_s}{m_s}Y(\Lambda_{s}r_{ij})\right],  \nonumber \\
& & s=\sigma,a_0,\kappa,f_0
\end{eqnarray}
where $T_{cm}$ is the kinetic energy of the center-of mass motion, $\mu$ is the reduced mass between two interacting quarks. And only the central parts
of the interactions are given here because we are interested in the low-lying states of the multiquark system. $\boldsymbol{\sigma}$
represents the SU(2) Pauli matrices; $\boldsymbol{\lambda}^c$, $\boldsymbol{\lambda}$ represent the SU(3) color and flavor Gell-Mann matrices
respectively; $\alpha_s$ denotes the strong coupling constant of one-gluon exchange and $Y(x)$ is the standard Yukawa functions.
Because it is difficult to have a good description of baryon and meson spectra simultaneously using the same set of parameters, we treat the strong
coupling constant of one-gluon exchange with different values for quark-quark and quark-antiquark interacting pairs. For the scalar nonet, we
use the same values for mass ($m_s$) and cut-off ($\Lambda_s$).

The model parameters are listed in Table \ref{parameter}, and the calculated baryon and meson masses are presented in the Table \ref{masss} with the 
experimental values.
From the calculation, the theoretical results of baryons contain a charm quark can fit well with the theoretical results. For meson spectrum,
most of the results were close to experimental values except for $\rho$ and $\omega$ mesons.
\begin{table}[h]
\caption{Quark model parameters.\label{parameter}}
\begin{tabular}{ccc} \hline
                  &$m_u$=$m_d$ (MeV)   & 313\\
Quark masses      &$m_s$ (MeV)  & 525\\
                  &$m_c$ (MeV)  & 1800\\   \hline
                   &$\Lambda_\pi$ (fm$^{-1}$)  & 4.20\\
                   &$\Lambda_\eta=\Lambda_K~$ (fm$^{-1}$)      & 5.20\\
                   &$m_\pi$ (fm$^{-1}$)  & 0.70\\
Goldstone bosons   &$m_K$ (fm$^{-1}$)  & 2.51\\
                   &$m_\eta$ (fm$^{-1}$)  & 2.77\\
                   &$g^2_{ch}/(4\pi)$  & 0.54\\
                   &$\theta_P(^\circ)$  & -15\\  \hline
                   &$a_c$ (MeV)  & 160.5\\
     Confinement       &$\mu_c$ (fm$^{-1})$  & 0.683\\
                    &$\Delta$ (MeV)  & 68.4\\   \hline
                   &$m_\sigma$ (fm$^{-1}$)  &   3.42\\
                   &$\Lambda_\sigma$ (fm$^{-1}$)  & 4.20\\
scalar nonet       &$\Lambda_{s}$ (fm$^{-1}$)  & 5.20\\
                   &$m_{s}$ (fm$^{-1}$)  & 4.97\\  \hline
                 &$\hat{r}_0~$(MeV~fm)  & 30.8\\
                    &$\alpha_{uu}$  & 0.552/0.684\\
        OGE           &$\alpha_{us}$  & 0.650/0.613\\
                    &$\alpha_{uc}$  &  0.633/0.683\\
                   &$\alpha_{sc}$  & 0.650/-\\   \hline
\end{tabular}
\end{table}

\begin{table}[h]
\caption{The masses of ground-state baryons and mesons (unit: MeV).\label{masss}}
\begin{tabular}{ccccccc}
\hline
 &$\Lambda$~ &$\Sigma$~ &$\Sigma^*$~ &$\Xi_c$~ &$\Xi_c'$~     \\ \hline
ChQM~ &1105~ &1201~ &1289~ &2512~ &2587~  \\
Exp.~\cite{1}~ &1116~ &1189~ &1385~ &2471~ &2579~  \\ \hline
 &$\Xi_c^*$~ &$\Sigma_c$~  &$\Sigma_c^*$~ &$\Lambda_c$~ \\ \hline
ChQM~ &2623~  &2449~ &2483~ &2297~  \\
Exp.~\cite{1}~  &2645~ &2455~ &2520~ &2286~  \\ \hline

&$\pi$~ &$\rho$~ &$\omega$~ &$\eta$~ &$K$~  \\ \hline
ChQM~ &140~ &698~ &607~ &542~ &494~  \\
Exp.~\cite{1}~ &140~ &775~ &782~ &548~ &494~  \\
 &$K^*$ &$D$~ &$D^*$  \\ \hline
ChQM~  &868~ &1864~ &2020~ \\
Exp.~\cite{1}~ &892~ &1864~ &2007~ \\
\hline
\end{tabular}
\end{table}

Since the newly observed states can be three-quark or five-quark states, a comprehensive calculation both involving three-quark and five-quark structure
is necessary. The calculation results of $\Xi_c^0$ ($usc$ in three-quark structure) for the low-lying excited ($2S$, $1P$) are listed in Table \ref{massp}.
From the results, we can see that the masses of all three $1P$ $\Xi_c^0$ states range from 2750 MeV to 2819 MeV, which are closed to the masses of two
exited $\Xi_c^0$ states with negative parity: $\Xi_c(2790)$ and $\Xi_c(2815)$. Similarly, the masses of $2S$ states of $\Xi_c'$ and $\Xi_c^*$ are closed to
the 2.9 GeV. Considering that our calculation result of $\Xi_c^*(2645)$ are lower than the experimental value of 22 MeV and if we make an energy correction, it will fit better with the experimental values. So we think that some of the three newly observed states can be the $2S$ states of $\Xi_c'$ and $\Xi_c^*$ if their parity are measured to be positive in the future experiments.
\begin{table}[h]
\caption{The masses of ground-state and low-lying exited $\Xi_c^0$ states (unit: Mev).\label{massp}}
\begin{tabular}{ccccc}
\hline
$nL$~  &$\Xi_c(2471)$~ &$\Xi_c'(2579)$~  &$\Xi_c^*$(2645)~   \\ \hline
$1S$~ &2512~ &2587~ &2623~   \\
$2S$~ &2824~ &2880~ &2900~   \\
$1P$~  &2750~ &2810~ &2819~  \\
\hline
\end{tabular}
\end{table}

Now, we turn to five-quark structure, and the same set of parameters in Table \ref{parameter} is used. It is worth mentioning that, the components of three-quark will
mix with that of five-quark and it will be necessary to consider a mixture of the two structures in future work.

In the following, we will construct the wave functions for the five-quark systems.
The wave function of the system consists of four parts: orbital, spin, flavor and color. The wave function of each part is constructed in two steps,
first construct the wave functions of three-quark cluster and quark-antiquark cluster, respectively, then coupling two clusters wave functions to form the
complete five-body one.

The first part is orbital wave function. A five-body system have four relative motions so it is written as follows,
\begin{equation}
\psi_{LM_L}=\left[ \left[ \left[
  \phi_{n_1l_1}(\mbox{\boldmath $\rho$})\phi_{n_2l_2}(\mbox{\boldmath $\lambda$})\right]_{l}
  \phi_{n_3l_3}(\mbox{\boldmath $r$}) \right]_{l^{\prime}}
  \phi_{n_4l_4}(\mbox{\boldmath $R$}) \right]_{LM_L},
\end{equation}
where the Jacobi coordinates are defined as follows,
\begin{eqnarray}
{\mbox{\boldmath $\rho$}} & = & {\mbox{\boldmath $x$}}_1-{\mbox{\boldmath $x$}}_2, \nonumber \\
{\mbox{\boldmath $\lambda$}} & = & (\frac{{m_1\mbox{\boldmath $x$}}_1+{m_2\mbox{\boldmath $x$}}_2}{m_1+m_2})-{\mbox{\boldmath $x$}}_3,  \nonumber \\
{\mbox{\boldmath $r$}} & = & {\mbox{\boldmath $x$}}_4-{\mbox{\boldmath $x$}}_5,  \\
{\mbox{\boldmath $R$}} & = & (\frac{{m_1\mbox{\boldmath $x$}}_1+{m_2\mbox{\boldmath $x$}}_2
  +{m_3\mbox{\boldmath $x$}}_3}{m_1+m_2+m_3})
  -(\frac{{m_4\mbox{\boldmath $x$}}_4+{m_5\mbox{\boldmath $x$}}_5}{m_4+m_5}). \nonumber
\end{eqnarray}
$\boldsymbol{x}_i$ is the position of the $i$-th particle. Then we use a set of gaussians to expand the radial part
of the orbital wave function which is shown below,
\begin{eqnarray}
\psi_{lm}(\mathbf{r})=\sum^{n_{max}}_{n=1}c_{nl}\phi^{G}_{nlm}(\mathbf{r})
\end{eqnarray}
\begin{eqnarray}
\phi^{G}_{nlm}(\mathbf{r})=\emph{N}_{nl}r^{l}e^{-\nu_{n}r^{2}}\emph{Y}_{lm}(\hat{\mathbf{r}})
\end{eqnarray}
where $N_{nl}$ is the normalization constant,
\begin{eqnarray}
\emph{N}_{nl}=\left(\frac{2^{l+2}(2\nu_{n})^{l+3/2}}{\sqrt\pi(2l+1)!!}\right)^{\frac{1}{2}},
\end{eqnarray}
and $c_{nl}$ is the variational parameter, which is determined by the dynamics of the system. The Gaussian size
parameters are chosen according to the following geometric progression:
\begin{eqnarray}
\nu_{n}=\frac{1}{r^{2}_{n}}, r_{n}=r_{min}a^{n-1}, a=\left(\frac{r_{max}}{r_{min}}\right)^{\frac{1}{n_{max}-1}},
\end{eqnarray}
where $n_{max}$ is the number of Gaussian functions, and $n_{max}$ is determined by the convergence of the results.
In the present calculation, $n_{max}=8$.

The spin wave functions of 3-quark and 2-quark clusters are written as follow,
\begin{eqnarray}
&& |B_{\frac12,\frac12}^{\sigma1}\rangle =\frac{1}{\sqrt{6}}
   (2\alpha\alpha\beta-\alpha\beta\alpha-\beta\alpha\alpha),~~ \nonumber \\
&& |B_{\frac12,\frac12}^{\sigma2}\rangle =\frac{1}{\sqrt{2}}
   (\alpha\beta\alpha-\beta\alpha\alpha), \nonumber \\
&& |B_{\frac12,-\frac12}^{\sigma1}\rangle =\frac{1}{\sqrt{6}}
   (\alpha\beta\beta+\beta\alpha\beta-2\beta\alpha\alpha),~~ \nonumber \\
&& |B_{\frac12,-\frac12}^{\sigma2}\rangle =\frac{1}{\sqrt{2}}
   (\alpha\beta\beta-\beta\alpha\beta),  \\
&& |B_{\frac32,\frac32}^{\sigma}\rangle =\alpha\alpha\alpha,
~|B_{\frac32,\frac12}^{\sigma}\rangle =\frac{1}{\sqrt{3}}
   (\alpha\alpha\beta+\alpha\beta\alpha+\beta\alpha\alpha), \nonumber \\
&& |B_{\frac32,-\frac12}^{\sigma}\rangle =\frac{1}{\sqrt{3}}
(\alpha\beta\beta+\beta\alpha\beta+\beta\beta\alpha),
~|B_{\frac32,-\frac32}^{\sigma}\rangle =\beta\beta\beta,  \nonumber \\
&& |M_{1,1}^{\sigma}\rangle =\alpha\alpha,
~|M_{1,0}^{\sigma}\rangle =\frac{1}{\sqrt{2}} (\alpha\beta+\beta\alpha),
~|M_{1,-1}^{\sigma}\rangle =\beta\beta, \nonumber \\
&& |M_{0,0}^{\sigma}\rangle =\frac{1}{\sqrt{2}} (\alpha\beta-\beta\alpha).\nonumber
\end{eqnarray}
Then, by coupling the spin wave functions of two sub-clusters using Clebsch-Gordan Coefficients, the total five-quark
spin wave function can be constructed.
\begin{eqnarray}
|\chi^{\sigma1}_{\frac12,\frac12} \rangle & = & |B_{\frac12,\frac12}^{\sigma1}\rangle |M_{0,0}^{\sigma}\rangle,
\nonumber \\
|\chi^{\sigma2}_{\frac12,\frac12} \rangle & = & |B_{\frac12,\frac12}^{\sigma2}\rangle |M_{0,0}^{\sigma}\rangle,
\nonumber \\
|\chi^{\sigma3}_{\frac12,\frac12} \rangle & = & -\sqrt{\frac{2}{3}}|B_{\frac12,-\frac12}^{\sigma1}\rangle |M_{1,1}^{\sigma}\rangle+\sqrt{\frac{1}{3}}|B_{\frac12,\frac12}^{\sigma1}\rangle |M_{1,0}^{\sigma}\rangle,
\nonumber \\
|\chi^{\sigma4}_{\frac12,\frac12} \rangle & = & -\sqrt{\frac{2}{3}}|B_{\frac12,-\frac12}^{\sigma2}\rangle |M_{1,1}^{\sigma}\rangle+\sqrt{\frac{1}{3}}|B_{\frac12,\frac12}^{\sigma2}\rangle |M_{1,0}^{\sigma}\rangle,
\nonumber \\
|\chi^{\sigma5}_{\frac12,\frac12} \rangle & = & \sqrt{\frac{1}{2}}|B_{\frac32,\frac32}^{\sigma}\rangle |M_{1,-1}^{\sigma}\rangle-\sqrt{\frac{1}{3}}|B_{\frac32,\frac12}^{\sigma}\rangle |M_{1,0}^{\sigma}\rangle
\nonumber \\
&& +\sqrt{\frac{1}{6}}|B_{\frac32,-\frac12}^{\sigma}\rangle |M_{1,1}^{\sigma}\rangle, \nonumber \\
|\chi^{\sigma6}_{\frac32,\frac32} \rangle & = & -|B_{\frac12,\frac12}^{\sigma1}\rangle |M_{1,1}^{\sigma}\rangle,  \\
|\chi^{\sigma7}_{\frac32,\frac32} \rangle & = & -|B_{\frac12,\frac12}^{\sigma2}\rangle |M_{1,1}^{\sigma}\rangle,
\nonumber \\
|\chi^{\sigma8}_{\frac32,\frac32} \rangle & = & |B_{\frac32,\frac32}^{\sigma}\rangle |M_{0,0}^{\sigma}\rangle,
\nonumber \\
|\chi^{\sigma9}_{\frac32,\frac32} \rangle & = & \sqrt{\frac{3}{5}}|B_{\frac32,\frac32}^{\sigma}\rangle |M_{1,0}^{\sigma}\rangle-\sqrt{\frac{2}{5}}|B_{\frac32,\frac12}^{\sigma}\rangle |M_{1,1}^{\sigma}\rangle,
\nonumber \\
|\chi^{\sigma10}_{\frac52,\frac52} \rangle & = & |B_{\frac32,\frac32}^{\sigma}\rangle |M_{1,1}^{\sigma}\rangle,  \nonumber
\end{eqnarray}

Similarly, we can construct the flavor wave function. We have three flavor configurations of the system:
$(qqs)(c \bar q)$, $(qqc)(s \bar q)$ and $(qsc)(q \bar q)$ ($q=u$ or $d$). The flavor wave functions of
three-quark and quark-antiquark subclusters can be written as follows,
\begin{eqnarray}
&& |B_{1,1}^{f1}\rangle = uuc, ~|B_{1,0}^{f1}\rangle = \frac{1}{\sqrt{2}}(udc+duc),
~|B_{1,-1}^{f1}\rangle = ddc, \nonumber \\
&& |B_{0,0}^{f1}\rangle = \frac{1}{\sqrt{2}}(udc-duc), \nonumber \\
&& |B_{1,1}^{f2}\rangle = uus, ~|B_{1,0}^{f2}\rangle = \frac{1}{\sqrt{2}}(uds+dus),
~|B_{1,-1}^{f2}\rangle = dds, \nonumber \\
&& |B_{0,0}^{f2}\rangle = \frac{1}{\sqrt{2}}(uds-dus),  \\
&& |B_{\frac12,\frac12}^{f3}\rangle = \frac{1}{\sqrt{2}}(usc+suc),
~|B_{\frac12,-\frac12}^{f3}\rangle = \frac{1}{\sqrt{2}}(dsc+sdc), \nonumber \\
&& |B_{\frac12,\frac12}^{f4}\rangle = \frac{1}{\sqrt{2}}(usc-suc),
~|B_{\frac12,-\frac12}^{f4}\rangle = \frac{1}{\sqrt{2}}(dsc-sdc), \nonumber \\
&& |M_{\frac12,\frac12}^{f1}\rangle = s \bar d ,
~|M_{\frac12,-\frac12}^{f1}\rangle = -s \bar u , \nonumber \\
&& |M_{\frac12,\frac12}^{f2}\rangle = c \bar d ,
~|M_{\frac12,-\frac12}^{f2}\rangle = -c \bar u , \nonumber \\
&& |M_{1,1}^{f3}\rangle = u \bar d ,
~|M_{1,0}^{f3}\rangle = \frac{1}{\sqrt{2}}(-u \bar u+d \bar d) ,
~|M_{1,-1}^{f3}\rangle = -d \bar u , \nonumber \\
&& |M_{0,0}^{f3}\rangle = \frac{1}{\sqrt{2}}(-u \bar u-d \bar d) , \nonumber
\end{eqnarray}
For the five-quark system under the present investigation, the possible isospin quantum numbers are $\frac12$
and $\frac32$.
The flavor wave functions of five-quark system with isospin $I=\frac12$ can be obtained as follows,
\begin{footnotesize}
\begin{eqnarray}
|\chi^{f1}_{\frac12,\frac12} \rangle & = & \sqrt{\frac{2}{3}}|B_{1,1}^{f1}\rangle |M_{\frac12,-\frac12}^{f1}\rangle-\sqrt{\frac{1}{3}}|B_{1,0}^{f1}\rangle |M_{\frac12,\frac12}^{f1}\rangle,
\nonumber \\
|\chi^{f2}_{\frac12,\frac12} \rangle & = & |B_{0,0}^{f1}\rangle |M_{\frac12,\frac12}^{f1}\rangle,\nonumber \\
|\chi^{f3}_{\frac12,\frac12} \rangle & = & \sqrt{\frac{2}{3}}|B_{1,1}^{f2}\rangle |M_{\frac12,-\frac12}^{f2}\rangle-\sqrt{\frac{1}{3}}|B_{1,0}^{f2}\rangle |M_{\frac12,\frac12}^{f2}\rangle,
\nonumber \\
|\chi^{f4}_{\frac12,\frac12} \rangle & = & |B_{0,0}^{f2}\rangle |M_{\frac12,\frac12}^{f2}\rangle, \\
|\chi^{f5}_{\frac12,\frac12} \rangle & = & \sqrt{\frac{1}{3}}|B_{\frac12,\frac12}^{f3}\rangle |M_{1,0}^{f3}\rangle-\sqrt{\frac{2}{3}}|B_{\frac12,-\frac12}^{f3}\rangle |M_{1,1}^{f3}\rangle,
\nonumber \\
|\chi^{f6}_{\frac12,\frac12} \rangle & = & \sqrt{\frac{1}{3}}|B_{\frac12,\frac12}^{f4}\rangle |M_{1,0}^{f3}\rangle-\sqrt{\frac{2}{3}}|B_{\frac12,-\frac12}^{f4}\rangle |M_{1,1}^{f3}\rangle,
\nonumber \\
|\chi^{f7}_{\frac12,\frac12} \rangle & = & |B_{\frac12,\frac12}^{f3}\rangle |M_{0,0}^{f3}\rangle,\nonumber \\
|\chi^{f8}_{\frac12,\frac12} \rangle & = & |B_{\frac12,\frac12}^{f4}\rangle |M_{0,0}^{f3}\rangle . \nonumber
\end{eqnarray}
\end{footnotesize}
The flavor wave functions of five-quark system with isospin $I=\frac32$ can be obtained in the following,
\begin{footnotesize}
\begin{eqnarray}
|\chi^{f9}_{\frac32,\frac32} \rangle & = & |B_{1,1}^{f1}\rangle |M_{\frac12,\frac12}^{f1}\rangle,
\nonumber \\
|\chi^{f10}_{\frac32,\frac32} \rangle & = & |B_{1,1}^{f2}\rangle |M_{\frac12,\frac12}^{f2}\rangle,  \\
|\chi^{f11}_{\frac32,\frac32} \rangle & = & |B_{\frac12,\frac12}^{f3}\rangle |M_{1,1}^{f3}\rangle,
\nonumber \\
|\chi^{f12}_{\frac32,\frac32} \rangle & = & |B_{\frac12,\frac12}^{f4}\rangle |M_{1,1}^{f3}\rangle .  \nonumber
\end{eqnarray}
\end{footnotesize}

The last part is the color wave function. We consider both two kinds of color structures,
color-singlet and color-octet. The total color wave functions are written directly:
\begin{eqnarray}
|\chi^{c1} \rangle & = & \frac{1}{\sqrt{18}}(rgb-rbg+gbr-grb+brg-bgr) \nonumber \\
  & & (\bar r r+\bar gg+\bar bb).  \\
|\chi^{c2} \rangle & = & \frac{1}{\sqrt{192}}[(rbg-gbr+brg-bgr)(2\bar b b-\bar r r-\bar g g) \nonumber \\
 & + & (2rgb-rbg+2grb-gbr-brg-bgr)(\bar r r-\bar g g) \nonumber \\
 & + & 2(2rrg-rgr-grr)\bar r b+2(rgg+grg-2ggr)\bar g b \nonumber \\
 & + & 2(2rrb-rbr-brr)\bar r g-2(rbb+brb-2bbr)\bar b g \nonumber \\
 & + & 2(2ggb-gbg-bgg)\bar g r+2(gbb+bgb-2bbg)\bar b r]. \nonumber \\
 & & \\
|\chi^{c3} \rangle & = & \frac{1}{\sqrt{576}}[3(rbg+gbr-brg-bgr)(\bar r r-\bar g g) \nonumber \\
& + & (2rgb+rbg-2grb-gbr-brg+bgr) \nonumber \\
&   & (2\bar b b-\bar r r-\bar g g) \nonumber \\
& + & 6(rgr-grr)\bar r b+6(rgg-grg)\bar g b \nonumber \\
& - & 6(rbr-brr)\bar r g-6(rbb-brb)\bar b g \nonumber \\
& + & 6(gbg-bgg)\bar g r+6(gbb-bgb)\bar b r].
\end{eqnarray}
Where $\chi^{c1} $ represents the color wave function of the color singlet structure. $\chi^{c2} $ and $\chi^{c3} $ represent the symmetric and
antisymmetric structures of the color-octet wave functions respectively, which their symmetry is between the first and second quark in baryon cluster.

Finally, the total wave function of the five-quark system is written as:
\begin{eqnarray}
&& \Psi_{JM_J}^{i,j,k}={\cal A} \left[ \left[
 \psi_{L}\chi^{\sigma_i}_{S}\right]_{JM_J}
   \chi^{fi}_j \chi^{ci}_k \right],\nonumber
\end{eqnarray}
where the $\cal{A}$ is the antisymmetry operator of the system which guarantees the antisymmetry of the total wave functions when identical
particles exchange.

The last step,we solve the following Schrodinger equation to obtain eigen-energies of the system.
\begin{equation}
H\Psi_{JM_J}=E\Psi_{JM_J},
\end{equation}
with the help of the Rayleigh-Ritz variational principle, the final result can be easily obtained if we just consider the situation five-quark systems
are all in ground states. It is worthwhile to mention if the orbital angular momenta of the system is not zero, it is necessary to use the infinitesimally
shifted Gaussian method~\cite{22}.

\section{Results and discussions}
\begin{figure}[h!]
  \centering
  \includegraphics[width=9cm,height=7cm]{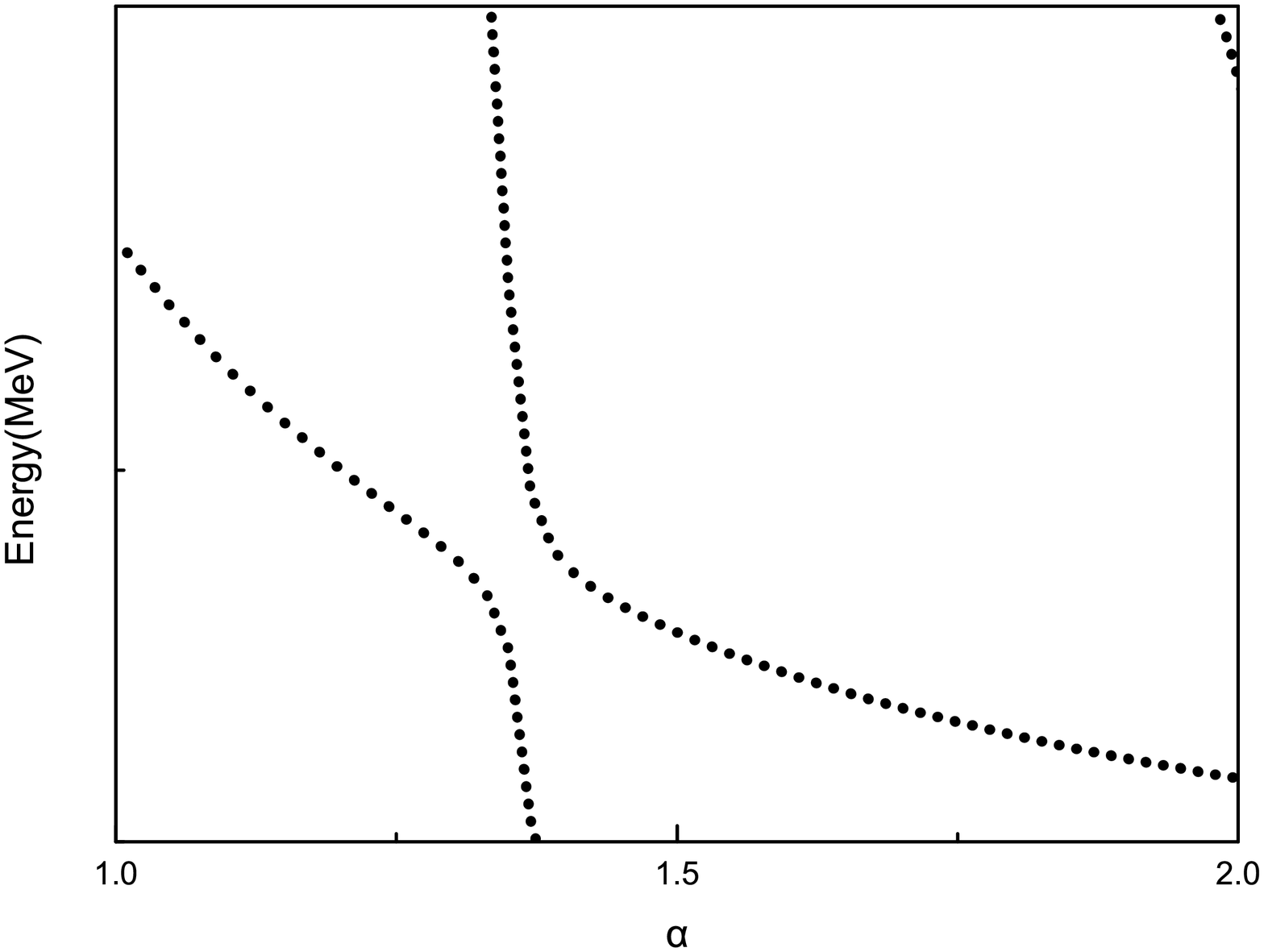}
  \caption{The shape of the resonance in real-scaling method.}
  \label{1}
\end{figure}

In this section, we will present calculation of all low-lying states of the $(usc)(q \bar q)$, $(uuc)(s \bar q)$ and $(uus)(c \bar q)$ pentaquark system
with all possible quantum numbers $IJ^P=\frac{1}{2}(\frac{1}{2})^-$, $\frac{1}{2}(\frac{3}{2})^-$, $\frac{1}{2}(\frac{5}{2})^-$, $\frac{3}{2}(\frac{1}{2})^-$, $\frac{3}{2}(\frac{3}{2})^-$ and $\frac{3}{2}(\frac{5}{2})^-$ by ChQM. All the orbital angular momenta of the system are set to
zero and so the corresponding parity is negative. To see if there exists genuine resonance states, we employ the real-scaling method~\cite{25,26,27,28} 
to make a check. In this method, the Gaussian size parameters $r_{n}$ for the basis
functions between baryon and meson clusters for the color-singlet channels are scaled by multiplying a factor $\alpha$: $r_{n} \longrightarrow \alpha r_{n}$.
As a result, the continuum state will fall off towards its threshold, and the resonant state would act with the scattering states and emerges as 
avoid-crossing structures with the variation of $\alpha$. A schematic diagram of a resonance is shown in Fig.~1. The top line is a scattering state, 
and it would fall to the corresponding threshold with increasing $\alpha$.
However, the down line, resonance state, would interact with the scattering state, which could result in a avoid-crossing structure.
If this behavior is repeated periodically, then the avoid-crossing point may be a stable resonance~\cite{29}.

What we consider important calculation results of are listed in Tables \ref{mass12} and \ref{mass32}. In each table, columns
2 to 5 represent the wave functions for the three degrees of freedom and their the physical channels of five-quark system. In column 6, eigen-energy of the each channel is shown and the theoretical value of noninteracting baryon-meson threshold (the sum of the masses of the corresponding baryon and meson in theory)
is along with it in column 7. Column 8 gives the binding energies, the difference between the eigen-energies and the theoretical thresholds. Finally, we will give the experimental thresholds (the sum of the experimental masses of the corresponding baryon and meson) along with corrected energies (the sum of experimental thresholds and the binding energies, $E^{\prime}=E_B+E_{th}^{Theo}$) in last two columns. We hope that this method can reduce the calculation error caused by the model parameters in five-quark calculation partly.
\begin{table*}[htb]
\caption{The eigen-energies of channels with $IJ^P=\frac{1}{2}\frac{1}{2}^-$. ``cc1" denotes the color-singlet channels coupling
and ``cc2" means full channels coupling (unit: MeV).\label{mass12}}
\begin{tabular}{cccccccccc}
\hline
~~Index~~&~~~~$\psi^{f_i}$~~~~&~~~~~$\psi^{\sigma_j}$~~~~~&~~~~~$\psi^{c_k}$~~~~~&~~~~Physical channel~~~~&~~$ E$ ~~& ~~$E^{Theo}_{th}$~~ &~~ $E_B$ ~~& ~~$E^{Exp}_{th}$ ~~&~~ $E^{\prime}$ ~~\\
\hline
1& $i=1$ & $j=1$ & $k=1$ & $\Sigma_c\bar{K}$ & 2944 & 2944  & 0 & 2949 & 2949 \\
2& $i=1$ & $j=1,2$ & $k=1,2,3$ &  & 2944 & \\
3& $i=1$ & $j=3$ & $k=1$ & $\Sigma_c\bar{K}^{*}$ & 3317 & 3317 & 0 & 3347 & 3347  \\
4& $i=1$ & $j=3,4$ & $k=1,2,3$ & & 3317 &  &  &  &  \\
5& $i=1$ & $j=5$ & $k=1$ & $\Sigma_c^*\bar{K}^{*}$ & 3351 & 3351 & 0 & 3412 & 3412  \\
6& $i=1$ & $j=5$ & $k=1,2,3$ & & 3351 &  &  &  &  \\
7& $i=2$ & $j=2$ & $k=1$ & $\Lambda_c\bar{K}$ & 2791 & 2791 & 0 & 2780 & 2780  \\
8& $i=2$ & $j=1,2$ & $k=1,2,3$ & & 2791 \\
9& $i=2$ & $j=2$ & $k=1$ & $\Lambda_c\bar{K}^{*}$ & 3165 & 3165 & 0 & 3178 & 3178  \\
10& $i=2$ & $j=1,2$ & $k=1,2,3$ & & 3165  \\
11& $i=3$ & $j=1$ & $k=1$ & $\Sigma D$ & 3062 & 3065  & -3 & 3053 & 3050 \\
12& $i=3$ & $j=1,2$ & $k=1,2,3$ &  & 3062 & & -3 & &3050 \\
13& $i=3$ & $j=3$ & $k=1$ & $\Sigma D^{*}$ & 3221 & 3221 & 0 & 3196 & 3196  \\
14& $i=3$ & $j=3,4$ & $k=1,2,3$ & & 3221  \\
15& $i=3$ & $j=5$ & $k=1$ & $\Sigma^*D^{*}$ & 3309 & 3309 & 0 & 3389 & 3389  \\
16& $i=3$ & $j=5$ & $k=1,2,3$ & & 3309 &  &  &  &  \\
17& $i=4$ & $j=2$ & $k=1$ & $\Lambda D$ & 2969 & 2969 & 0 & 2980 & 2980  \\
18& $i=4$ & $j=1,2$ & $k=1,2,3$ & & 2969 \\
19& $i=4$ & $j=2$ & $k=1$ & $\Lambda D^{*}$ & 3125 & 3125 & 0 & 3123 & 3123  \\
20& $i=4$ & $j=1,2$ & $k=1,2,3$ & & 3125 & &  & &   \\
21& $i=5$ & $j=1$ & $k=1$ & $\Xi_c\pi$ & 2652 & 2652  & 0 & 2611 & 2611 \\
22& $i=5$ & $j=1,2$ & $k=1,2,3$ &  & 2652 & \\
23& $i=6$ & $j=2$ & $k=1$ & $\Xi_c'\pi$ & 2727 & 2727 & 0 & 2719 & 2719  \\
24& $i=6$ & $j=1,2$ & $k=1,2,3$ & & 2727  \\
25& $i=5$ & $j=3$ & $k=1$ & $\Xi_c\rho$ & 3210 & 3210 & 0 & 3245 & 3245  \\
26& $i=5$ & $j=3,4$ & $k=1,2,3$ & & 3210  \\
27& $i=6$ & $j=4$ & $k=1$ & $\Xi_c'\rho$ & 3285 & 3285 & 0 & 3354 & 3354  \\
28& $i=6$ & $j=3,4$ & $k=1,2,3$ & & 3285 \\
29& $i=7$ & $j=1$ & $k=1$ & $\Xi_c\eta$ & 3054 & 3054 & 0 & 3019 & 3019  \\
30& $i=7$ & $j=1,2$ & $k=1,2,3$ & & 3054  \\
31& $i=8$ & $j=2$ & $k=1$ & $\Xi_c'\eta$ & 3129 & 3129  & 0 & 3127 & 3127 \\
32& $i=8$ & $j=1,2$ & $k=1,2,3$ &  & 3129  \\
33& $i=7$ & $j=3$ & $k=1$ & $\Xi_c\omega$ & 3119 & 3119 & 0 & 3258 & 3258  \\
34& $i=7$ & $j=3,4$ & $k=1,2,3$ & & 3119  \\
35& $i=8$ & $j=4$ & $k=1$ & $\Xi_c'\omega$ & 3194 & 3194 & 0 & 3361 & 3361  \\
36& $i=8$ & $j=3,4$ & $k=1,2,3$ & & 3194 \\
37& $i=5$ & $j=5$ & $k=1$ & $\Xi_c^*\rho$ & 3321 & 3321 & 0 & 3423 & 3423  \\
38& $i=5$ & $j=5$ & $k=1,2,3$ & & 3321 \\
39& $i=7$ & $j=5$ & $k=1$ & $\Xi_c^*\omega$ & 3230 & 3230 & 0 & 3427 & 3427  \\
40& $i=7$ & $j=5$ & $k=1,2,3$ & & 3230  \\
cc1 & & & & & 2652 &  & 0  \\
cc2 & & & & & 2652  \\
\hline
\end{tabular}
\end{table*}
\begin{table*}[htb]
\caption{The eigen-energies of channels with $IJ^P=\frac{1}{2}\frac{3}{2}^-$ ``cc1" denotes the color-singlet channels coupling
and ``cc2" means full channels coupling (unit: MeV).\label{mass32}}
\begin{tabular}{cccccccccc}
\hline 
~~Index~~&~~~~$\psi^{f_i}$~~~~&~~~~~$\psi^{\sigma_j}$~~~~~&~~~~~$\psi^{c_k}$~~~~~&~~~~Physical channel~~~~&~~$ E$ ~~& ~~$E^{Theo}_{th}$~~ &~~ $E_B$ ~~& ~~$E^{Exp}_{th}$ ~~&~~ $E^{\prime}$ ~~\\
\hline
1& $i=1$ & $j=6$ & $k=1$ & $\Sigma_c\bar{K}^{*}$ & 3317 & 3317  & 0 & 3344 & 3344 \\
2& $i=1$ & $j=6,7$ & $k=1,2,3$ &  & 3317   \\
3& $i=1$ & $j=8$ & $k=1$ & $\Sigma_c^*\bar{K}$ & 2978 & 2978 & 0 & 3014 & 3014  \\
4& $i=1$ & $j=8$ & $k=1,2,3$ & & 2978 &  &  &  &  \\
5& $i=1$ & $j=9$ & $k=1$ & $\Sigma_c^*\bar{K}{*}$ & 3350 & 3350 & 0 & 3412 & 3412  \\
6& $i=1$ & $j=9$ & $k=1,2,3$ & & 3350 &  &  &  &  \\
7& $i=2$ & $j=7$ & $k=1$ & $\Lambda_c\bar{K}^*$ & 3165 & 3165 & 0 & 3178 & 3178  \\
8& $i=2$ & $j=6,7$ & $k=1,2,3$ & & 3165 &  &  &  &  \\
9& $i=3$ & $j=6$ & $k=1$ & $\Sigma D^*$ & 3217 & 3221 & -4 & 3196 & 3192  \\
10& $i=3$ & $j=6,7$ & $k=1,2,3$ & & 3215 & & -6 & & 3190  \\
11& $i=3$ & $j=8$ & $k=1$ & $\Sigma^* D$ & 3152 & 3153 & -1 & 3246 & 3245  \\
12& $i=3$ & $j=8$ & $k=1,2,3$ & & 3151 & & -2 & & 3244 \\
13& $i=3$ & $j=9$ & $k=1$ & $\Sigma^*D^*$ & 3305 & 3309 & -4 & 3389 & 3385  \\
14& $i=3$ & $j=6,7$ & $k=1,2,3$ & & 3304 & & -5 & & 3384 \\
15& $i=4$ & $j=7$ & $k=1$ & $\Lambda D^*$ & 3125 & 3125 & 0 & 3123 & 3123  \\
16& $i=4$ & $j=6,7$ & $k=1,2,3$ & & 3125 & &  & &  \\
17& $i=5$ & $j=6$ & $k=1$ & $\Xi_c\rho$ & 3210 & 3210 & 0 & 3245 & 3245  \\
18& $i=5$ & $j=6,7$ & $k=1,2,3$ & & 3210  \\
19& $i=5$ & $j=8$ & $k=1$ & $\Xi_c^*\pi$ & 2763 & 2763  & 0 & 2785 & 2785 \\
20& $i=5$ & $j=8$ & $k=1,2,3$ &  & 2763 & \\
21& $i=5$ & $j=9$ & $k=1$ & $\Xi_c^*\rho$ & 3321 & 3321 & 0 & 3423 & 3423  \\
22& $i=5$ & $j=9$ & $k=1,2,3$ & & 3321  \\
23& $i=6$ & $j=7$ & $k=1$ & $\Xi_c'\rho$ & 3285 & 3285 & 0 & 3354 & 3354  \\
24& $i=6$ & $j=6,7$ & $k=1,2,3$ & & 3285  \\
25& $i=7$ & $j=6$ & $k=1$ & $\Xi_c\omega$ & 3119 & 3119 & 0 & 3259 & 3259  \\
26& $i=7$ & $j=6,7$ & $k=1,2,3$ & & 3119  \\
27& $i=7$ & $j=8$ & $k=1$ & $\Xi_c^*\eta$ & 3165 & 3165  & 0 & 3193 & 3193 \\
28& $i=7$ & $j=8$ & $k=1,2,3$ &  & 3165 & \\
29& $i=7$ & $j=9$ & $k=1$ & $\Xi_c^*\omega$ & 3230 & 3230 & 0 & 3427 & 3427  \\
30& $i=7$ & $j=9$ & $k=1,2,3$ & & 3230  \\
31& $i=8$ & $j=7$ & $k=1$ & $\Xi_c'\omega$ & 3194 & 3194 & 0 & 3361 & 3361  \\
32& $i=8$ & $j=6,7$ & $k=1,2,3$ & & 3194  \\
cc1 & & & & & 2763 &  & 0  \\
cc2 & & & & & 2763  \\
\hline 
\end{tabular}
\end{table*}

Since the energy of the three newly observed states is between 2.9 GeV and 3.0 GeV, for the systems with $I=\frac{3}{2}$, there is hardly any bound state,
and even if there is bound state, the energy is much higher than 3.0 GeV. Also for the system with $IJ^P=\frac{1}{2}\frac{5}{2}^-$, the energy of all
channels are all around 3.4 GeV. So we focused our analysis on systems with $IJ^P=\frac{1}{2}\frac{1}{2}^-$ and $IJ^P=\frac{1}{2}\frac{3}{2}^-$.
We analyze the results of these systems in detail in the following:

(a) $IJ^P=\frac{1}{2}\frac{1}{2}^-$: The single channel calculations show that there are weak attractions for the $\Sigma D$ channel which is consistent
with conclusions that there are attractions between $D$ meson and $\Sigma$  baryon in Refs.~\cite{16,18,19}. After coupling to the respective hidden-color
channels, our results did not change. Then, we make multi-channel coupling and the calculations show that no bound state can be formed. 

(b) $IJ^P=\frac{1}{2}\frac{3}{2}^-$: We have the similar results with the case of $IJ^P=\frac{1}{2}\frac{1}{2}^-$. $\Sigma D^*$ channels have weak
attractions and after coupling to respective hidden-color channels, most single channels ($\Sigma^*D^*$, $\Sigma^*D$ and $\Sigma^*D^*$) exist weak
attractions except for $(usc)(q \bar q)$ and $(uuc)(s \bar q)$ systems. No bound state can be formed in this system after multi-channel coupling
and it is necessary to do a resonance state test by real-scaling method.

(c) For $IJ^P=\frac{1}{2}\frac{5}{2}^-$ ,$\frac{3}{2}(\frac{1}{2})^-$, $\frac{3}{2}(\frac{3}{2})^-$ and $\frac{3}{2}(\frac{5}{2})^-$ systems, there is
no bound state and their threshold is much more higher than 3.0 GeV, so we do not put the calculation up.

The results of $IJ^P=\frac{1}{2}\frac{1}{2}^-$ and $\frac{1}{2}\frac{3}{2}^-$ with help of real-scaling method are shown in Figs. 2 and 3.
In these four figures, we have marked the threshold with lines (red online) and the physical components are also tagged. The mass for $2S$ states of $\Xi_c$,
$\Xi_c'$ and $\Xi_c^*$ are given in Table III. For the possible resonances, we mark it in line with energy as a tag (blue online). Since we focus on 
the energy around 3.0 GeV,
we set the energy range between 2650 MeV and 3080 MeV. From the figures, we can see that all thresholds appear as horizontal lines. In addition,
we can find genuine resonances whose energy are stable with the increase of $\alpha$. For $IJ^P=\frac{1}{2}\frac{1}{2}^-$ system, we got two resonances that energy are 2975 MeV  and 3029 MeV around 3.0 GeV. In the same way, we also find one resonance in $IJ^P=\frac{1}{2}\frac{3}{2}^-$ system : 3048 MeV. Because of too many channels involved, We did a simple energy correction to these resonances and the result is 2954 MeV($\Sigma D$), 3017 MeV($\Sigma D$) and 3020 MeV($\Sigma D^{*}$). Considering the deviations of quark model calculations and the energy of the resonances are very closed to the newly observed states, these two resonances are good candidates of the newly reported states.
\begin{figure}[h!]
  \centering
  \includegraphics[width=9cm,height=7cm]{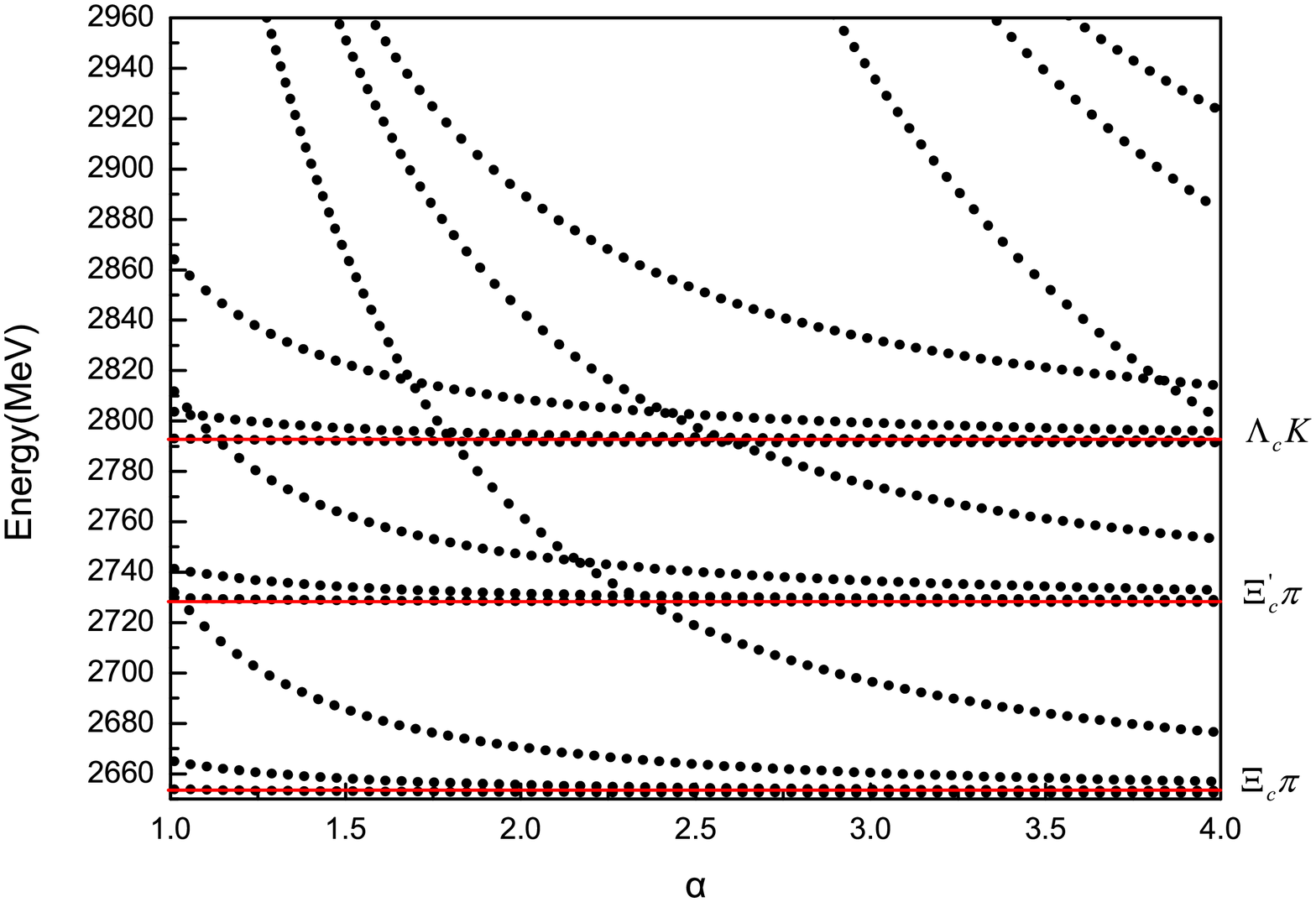}
  \includegraphics[width=9cm,height=7cm]{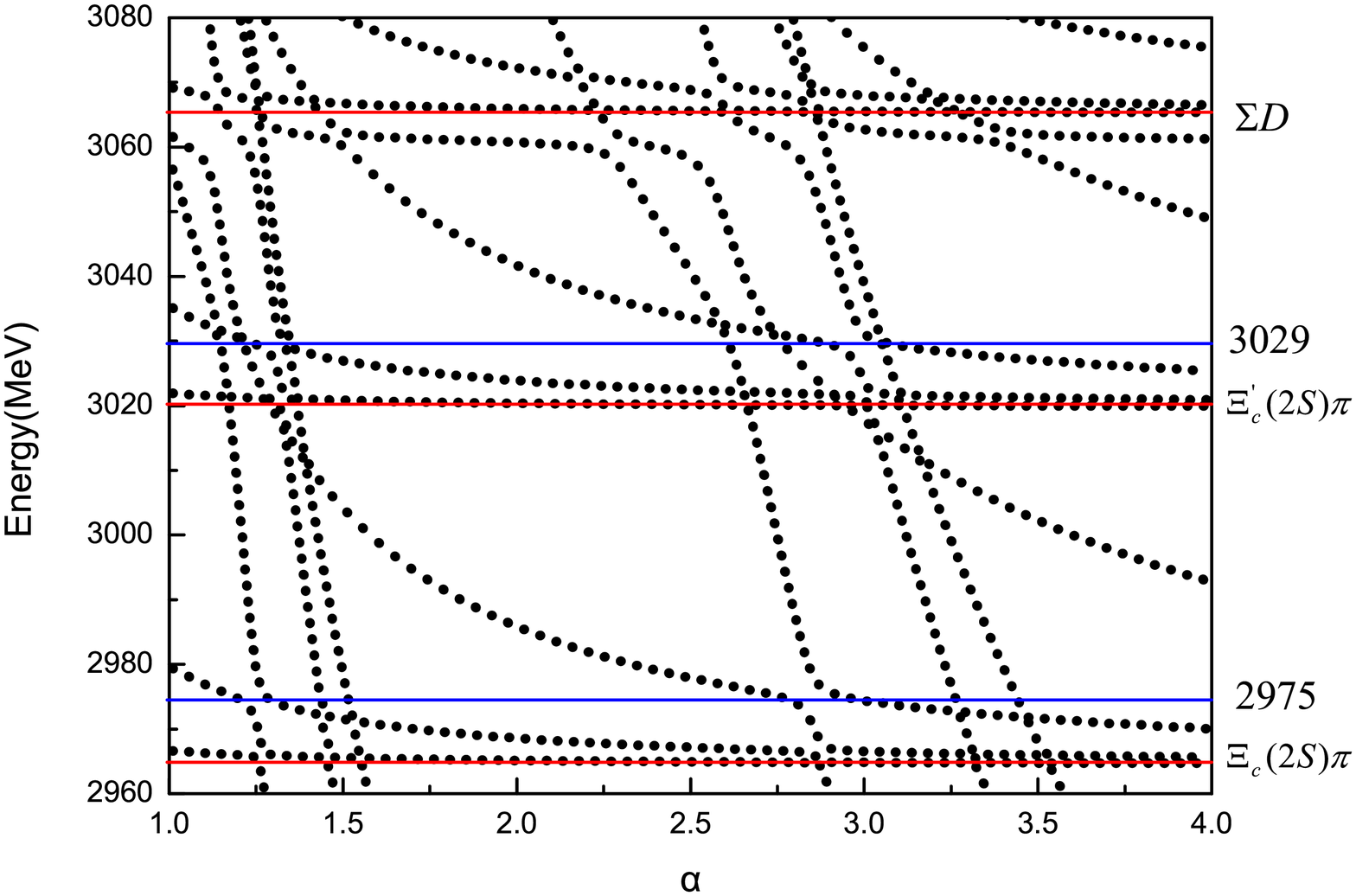}
  \caption{Energy spectrum of $\frac{1}{2}\frac{1}{2}^-$ system.}
  \label{2}
\end{figure}
\begin{figure}[h!]
  \centering
  \includegraphics[width=9cm,height=7cm]{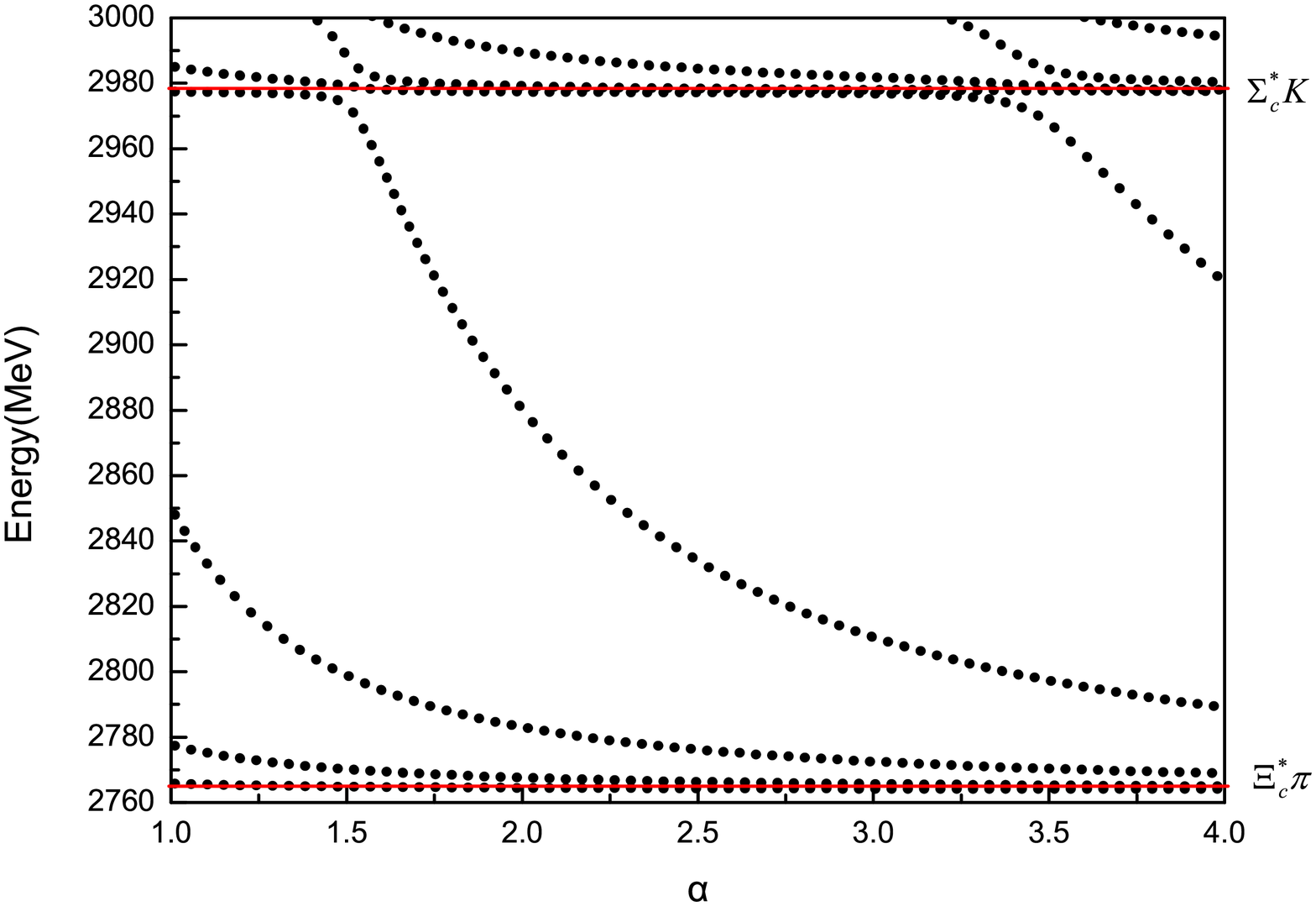}
  \includegraphics[width=9cm,height=7cm]{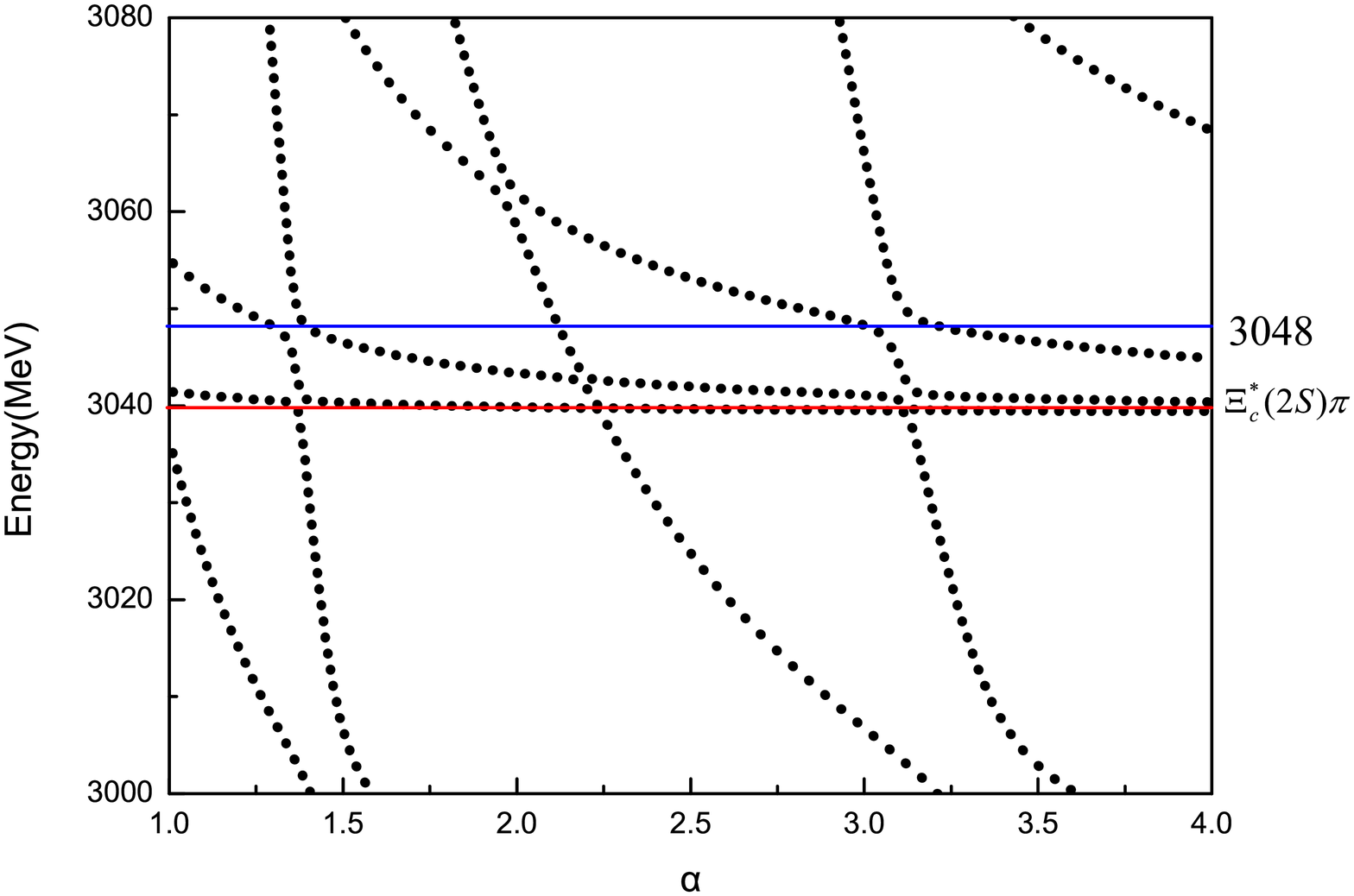}
  \caption{Energy spectrum of $\frac{1}{2}\frac{3}{2}^-$ system.}
  \label{3}
\end{figure}

We also do a further calculation about partial decay width of these two resonances. The decay widths for three resonances to possible open channels
are obtained by the following formula:
\begin{equation}
\Gamma=4V(\alpha)\frac{\sqrt{k_r  k_c}}{|k_r-k_c|},
\end{equation}
where $V(\alpha)$ is the minimum energy difference, while $k_c$ and $k_r$ stands for the slopes of scattering state and resonance state at the
avoid-crossing point respectively. More details can be found in Ref.~\cite{30}. The results of decay width are shown in Table \ref{width}.
\begin{table}[h]
\caption{The decay width of $\Xi_c^0$ states with mass of 2975 MeV, 3029 MeV and 3048 MeV (unit: MeV). \label{width}}
\begin{tabular}{ccccccc}
\hline 
$IJ^P$  &State &Width & $E^{\prime}$ & Candidate   \\ \hline
$\frac{1}{2}(\frac{1}{2})^-$~ &$\Xi_c(2975)$~ &8.2~ &2954~ & $\Xi_c(2923)$  ~\\\hline
$\frac{1}{2}(\frac{1}{2})^-$~ &$\Xi_c(3029)$~ &10.1~ &3017~ & $\Xi_c(2938)$  ~\\\hline
$\frac{1}{2}(\frac{3}{2})^-$~ &$\Xi_c(3048)$~ &12.9~ &3020~ & $\Xi_c(2965)$~  \\
\hline 
\end{tabular}
\end{table}

We can see that for $\Xi_c(2975)$, $\Xi_c(3029)$ and $\Xi_c(3048)$ states, the decay width are 8.2 MeV, 10.1 MeV and 12.9 MeV respectively. After considering the statistic uncertainty and systematic uncertainty of experimental results, we think these three resonances can be good candidates for $\Xi_c(2923)$, $\Xi_c(2938)$ and $\Xi_c(2965)$ if the parity of them are found to be negative.

\section{Summary}
In this paper, we investigate the three-quark $usc$ states and five-quark $uscu\bar{u}$ or $uscd\bar{d}$ states in the framework of the chiral
constituent quark model with the help of Gaussian expansion method. For the result of three-quark system, we find that both first radial excitation
states ($2S$) of $\Xi_c'$(2579) and $\Xi_c^*$(2645) have the energy closed to 2.9 GeV, which are possible candidates of the newly observed excited states
of $\Xi_c$ reported by LHCb Collaboration. It is worth mentioning that our preliminary calculation indicated the energy of $1P$ states of $\Xi_c$ are around
2.8 GeV, which are good candidates of $\Xi_c^*$(2790) and $\Xi_c^*$(2815) from PDG with consistent negative-parity. For the five-quark systems, we investigate
all the possible systems with negative-parity, especially systems with $IJ^P=\frac{1}{2}\frac{1}{2}^-$ and $IJ^P=\frac{1}{2}\frac{3}{2}^-$ and the calculation
shows that there is no bound state for these systems. However, the resonances are possible. We use real-scaling method to find the genuine resonances and
also study their decay width. Three resonance states $\Xi_c(2975)$, $\Xi_c(3029)$ and $\Xi_c(3048)$ with energy around 2.9 GeV to 3.0 GeV have been found in the $IJ^P=\frac{1}{2}\frac{1}{2}^-$ and $IJ^P=\frac{1}{2}\frac{3}{2}^-$ systems. These three resonance states are possible candidates of the newly reported excited states of $\Xi_c$ by LHCb Collaboration.

However, we cannot jump to conclusions with calculations of three-quark system and five-quark system in this paper because experimental and theoretical
studies of these three newly observed states is not enough. In addition, a mixed system of three-quark baryon and pentaquark cannot be ignored.
So, study of $\Xi_c$ in the framework of the unquenched quark model, including the higher Fock components with more future experimental and theoretical
data is our future work.

\section*{Acknowledgments}
The work is supported partly by the National Natural Science Foundation of China under Grant
Nos. 11775118 and 11535005.

\end{document}